# Morlet wavelets in quantum mechanics

John Ashmead

## *Abstract*

*Wavelets offer significant advantages for the analysis of problems in quantum mechanics. Because wavelets are localized in both time and frequency they avoid certain subtle but potentially fatal conceptual errors that can result from the use of plane wave or δ function decomposition. Morlet wavelets are particularly well-suited for this work: as Gaussians, they have a simple analytic form and they work well with Feynman path integrals. To take full advantage of Morlet wavelets we need an explicit form for the inverse Morlet transform and a manifestly covariant form for the four-dimensional Morlet wavelet. We supply both here.*

*In theory, theory and practice are the same. In practice they are not. – A. Einstein*

## Introduction

Wavelet transforms represent a natural development of Fourier transforms and may be used for similar purposes. Where the Fourier transform lets us decompose a wave function into its component plane waves, a wavelet transform lets us decompose a wave function into its component wavelets. If we think of the plane waves as corresponding to pure tones, we may think of the wavelets as corresponding to the notes produced by physical instruments: of finite duration and spanning a finite range of tones.

Wavelets have two advantages over plane waves:

1. They are localized in time and frequency. This can make them a better fit to the wave forms found in nature, which are always localized in both time and frequency. As a result, wavelet series will often converge faster than corresponding Fourier series.
2. There are many different wavelets to choose from: we can tailor our wavelets to our problem.

These advantages have resulted in their application to a wide variety of practical problems in acoustics, astronomy, medical imaging, computer graphics, meteorology, and so on[1].

Wavelets also have significant – if less numerous – application on the theory side. Examples: canonical quantization of the electromagnetic field using a discrete wavelet basis (Havukainen 2006), analysis of localization properties of photons using windowed wavelets (Kim 1996), regularization of Euclidean field theories (Altaisky 2003), and use of wavelets to provide "Lorentz covariant, singularity-free, finite energy, zero action, localized solutions to the wave equation" (Visser 2003).

Wavelets offer significant benefits for the study of foundational questions in quantum mechanics as well. We will focus here specifically on Morlet wavelets. These are Gaussians, so are both easy to work with and a natural fit to path integrals, which typically consist of long series of Gaussian integrations. Use of Morlet wavelets can let us:

1. Avoid any need to invoke the notorious "collapse of the wave function" in the analysis of the Stern-Gerlach experiment,
2. Avoid the use of artificial convergence factors or Wick rotation in computing path integrals,
3. Compute path integrals in a time symmetric way.

But to prepare Morlet wavelets for their new responsibilities we need to:

1. Supply an explicit form for the "admissibility constant" discussed below. This is needed to define the inverse Morlet transform.
2. Provide a manifestly covariant extension of Morlet wavelets to four dimensions.

[1] A search on the preprint archive arXiv.org turned up over one thousand preprints with wavelet in title or abstract.

Morlet's original reference is (Morlet, Arens et al. 1982). Wavelets are discussed in: (Chui 1992; Meyer 1992; Kaiser 1994; van den Berg 1999; Addison 2002; Bratteli and Jørgensen 2002; Antoine, Murenzi et al. 2004). Path integrals are discussed in: (Feynman and Hibbs 1965; Schulman 1981; Swanson 1992; Khandekar, Lawande et al. 1993; Marchewka and Schuss 2000; Kleinert 2004; Zinn-Justin 2005).

## Measurements, path integrals, and time

### *Stern-Gerlach experiment*

In the original Stern-Gerlach experiment (Gerlach and Stern 1922; Gerlach and Stern 1922; Gerlach and Stern 1922) a beam of silver atoms is sent through an inhomogeneous magnetic field. The beam is split into two: those atoms with spin up getting a kick in one direction; those with spin down in the opposite. This was striking both because it demonstrated the existence of spin and because a classical system would have shown a continuous range of values for the spin, not just up and down. This split is regarded as a classic demonstration of the measurement problem, explained – in the Copenhagen interpretation (von Neumann 1955) – by the “collapse of the wave function” into up and down components, and gotten much attention since.

Typically the initial wave function is modeled as a plane wave times a spin vector. But recently Gondran and Gondran (Gondran and Gondran 2005) have shown that if you model the initial wave function of the silver atom as a Gaussian, and turn the crank on the Schrödinger equation, you see the spin up and spin down components separate without any need to invoke a “collapse”. It works a bit like a diffraction experiment: there is coherent interference at two spots, incoherent at the rest. One may think of this as an internal diffraction effect.

Gondran and Gondran intended their work at least partly in support of the Bohm interpretation, however the math is independent of the interpretation. Standard quantum mechanics – if done in sufficient detail, i.e. with Gaussian test functions rather than plane waves – can explain the Stern-Gerlach effect without further assistance.

The use of a single Gaussian test function is not of itself general. But with the use of the Morlet wavelet transform we can write an arbitrary square-integrable wave function as a sum over Gaussian test functions, making the Gondran and Gondran result completely general.

To be sure, we could attempt to restore the honor of the plane wave by arguing that we could build up a Gaussian test function as a sum over such. But then why not “eliminate the middleman” and start with Gaussian test functions?

I am aware of several related examinations of the Stern-Gerlach effect[2]. Cruz-Barrios and Gómez-Camacho (Cruz-Barrios and Gómez-Camacho 1998; Cruz-Barrios and Gómez-Camacho 2001) argued that if the atoms could be modeled using “coherent internal states” (CIS), we would see the effect. And Venugopalan et al (Venugopalan, Kumar et al. 1995; Venugopalan 1997; Venugopalan 1999) argued we would see the effect as an effect of decoherence. The Gondran and Gondran result is simpler in that it posits no additional structure (CIS) or additional interaction (decoherence); standard quantum mechanics of its own gives the effect.

### *Convergence of path integrals*

Morlet wavelets can assist in establishing convergence of Feynman path integrals without recourse to convergence factors as used in (Feynman and Hibbs 1965; Schulman 1981) or Wick rotation as in (Zinn-Justin 2005); convergence of the slice-by-slice integrals in the path integral is a side-effect of the initial wave function being a (sum of) Gaussians, for which convergence is automatic. Step by step in the free case:

1. We start with the free Schrödinger equation:

$$i\frac{d}{d\tau}\psi_\tau(\vec{x}) = -\frac{1}{2m}\nabla^2\psi_\tau(\vec{x})$$

2. The path integral expression for the kernel is given by:

$$K_\tau(\vec{x};\vec{x}') = \lim_{N\to\infty}\sqrt{\frac{m}{2\pi i\hbar\varepsilon}}^{3N}\int \mathrm{d}\vec{x}_1\ldots\mathrm{d}\vec{x}_{N-1}\exp\left(\frac{i\varepsilon}{\hbar}\sum_{j=0}^{N-1}\frac{m}{2}\left(\frac{\vec{x}_{j+1}-\vec{x}_j}{\varepsilon}\right)^2\right)$$

3. A typical integral is:

$$K_j(\vec{x}_{j+1};\vec{x}_{j-1}) = \sqrt{\frac{m}{2\pi i\hbar\varepsilon}}^{3}\int \mathrm{d}\vec{x}_j\exp\left(\frac{i\varepsilon}{\hbar}\left(\frac{m}{2}\left(\frac{\vec{x}_{j+1}-\vec{x}_j}{\varepsilon}\right)^2+\frac{m}{2}\left(\frac{\vec{x}_j-\vec{x}_{j-1}}{\varepsilon}\right)^2\right)\right)$$

---

[2] See also Ashmead, J. (2003, Tue, 7 Jan 2003 08:30:07 GMT). "Uncollapsing the wave function." from http://lanl.arxiv.org/pdf/quant-ph/0301016.

Where ε is the width of a time slice:

$$\varepsilon \equiv \frac{\tau}{N}$$

4. The typical integral does not converge. We can force convergence by adding a small imaginary part $i\sigma$ to the mass:

   $$m \to m + i\sigma .$$

   We could also add the small imaginary part to the time step ε or Wick rotate into imaginary time:

   $$t \to it .$$

5. Now, focus attention on the first step:

   $$\psi_1(\vec{x}_1) = \int \mathrm{d}\vec{x}_0 K_1(\vec{x}_1;\vec{x}_0)\psi_0(\vec{x}_0)$$

6. Assume the initial wave function is a Gaussian:

   $$\psi_1(\vec{x}_1) = \sqrt{\frac{m}{2\pi i\hbar\varepsilon}}^3 \int \mathrm{d}\vec{x}_0 \exp\left(\frac{i\varepsilon}{\hbar}\left(\frac{m}{2}\left(\frac{\vec{x}_1 - \vec{x}_0}{\varepsilon}\right)^2\right)\right)\sqrt[4]{\frac{1}{\pi\sigma^6}} \exp\left(-\frac{\left(\vec{x}_0 - \langle\vec{x}_0\rangle\right)^2}{2\sigma^2}\right)$$

   This integral is convergent of itself. The result is a (slightly wider) Gaussian. We can do an infinite series of these, with the initial wave function showing an increasing amount of middle-aged spread but with all of the integrals converging.

   As an arbitrary wave function may be written, via Morlet wavelets, as a sum over Gaussian test functions, we have convergence in the general case[3], without the introduction of artificial convergence factors.

This by no means eliminates all of the technical problems with path integrals; for instance, there is still the curious question of the mid-point rule, as discussed in (Schulman 1981). But one step at a time.

### *Symmetric analysis of time in path integrals*

One immediate benefit of not needing convergence factors or Wick rotation is that we can treat time in a more symmetric way. One case where we might want to do this is in setting up a path integral analysis of the Stückelberg-Schrödinger equation:

$$i\frac{d\psi_u(x)}{du} = H\psi_u(x)$$

Here *u* is a formal parameter – a scalar of some kind, perhaps the particle's proper time -- and *H* is a Lorentz invariant Hamiltonian. There are examples in Feynman (Feynman 1950; Feynman 1951) and more recently in work by Land, Horwitz, and Seidewitz (Land and Horwitz 1996; Horwitz 1998; Seidewitz 2005). This approach has been sufficiently interesting that there is an ongoing conference devoted to this & related questions: (Gill, Horwitz et al. 2010).

In the free case *H* might be given by:

$$H = -\frac{1}{2m}\left(i\frac{\partial}{\partial x^\mu}\right)\left(i\frac{\partial}{\partial x_\mu}\right) = \frac{1}{2m}\left(\frac{\partial^2}{\partial t^2} - \frac{\partial^2}{\partial x^2} - \frac{\partial^2}{\partial y^2} - \frac{\partial^2}{\partial z^2}\right)$$

Note that because of the Lorentz invariance the time and space parts enter into *H* with opposite sign, so in the path integral will have a problem converging in a Lorentz covariant way:

1. Path integral form for the kernel:

   $$K_\tau(x;x') = \lim_{N\to\infty}\sqrt{\frac{m}{2\pi\hbar\varepsilon}}^{4N} \int \mathrm{d}t_1\,\mathrm{d}\vec{x}_1 \ldots \mathrm{d}t_{N-1}\,\mathrm{d}\vec{x}_{N-1} \exp\left(-\frac{i\varepsilon}{\hbar}\sum_{j=0}^{N-1}\frac{m}{2}\left(\left(\frac{t_{j+1} - t_j}{\varepsilon}\right)^2 - \left(\frac{\vec{x}_{j+1} - \vec{x}_j}{\varepsilon}\right)^2\right)\right)$$

   Note the pre-factor from the time part is slightly different from that for space:

[3] I have elided some technical difficulties here; discussed at (much) greater length in a work in progress Ashmead, J. (2009). Quantum Time. Philadelphia.

$$\sqrt{\frac{m}{2\pi i\hbar\varepsilon}} \to \sqrt{\frac{im}{2\pi\hbar\varepsilon}}$$

2. A typical slice:

$$K_j\left(\vec{x}_{j+1};\vec{x}_{j-1}\right) = \sqrt{\frac{m}{2\pi\hbar\varepsilon}}^4 \int \mathrm{d}\vec{x}_j \exp\left(-\frac{i\varepsilon}{\hbar}\frac{m}{2}\left(\left(\frac{t_{j+1}-t_j}{\varepsilon}\right)^2 - \left(\frac{\vec{x}_{j+1}-\vec{x}_j}{\varepsilon}\right)^2 + \left(\frac{t_j-t_{j-1}}{\varepsilon}\right)^2 - \left(\frac{\vec{x}_j-\vec{x}_{j-1}}{\varepsilon}\right)^2\right)\right)$$

3. But now the addition of a small imaginary part to mass or time fails; any change that causes the time integrals to converge will cause the space integrals to diverge and vice versa. If we use different signs for time and space, then we break covariance.
4. Wick rotation fails for the same reason. No matter which sign we choose for the Wick rotation, either the past or the future side will blow up.
5. Again, look at the first step:

$$\psi_1\left(t_1,\vec{x}_1\right) = \int \mathrm{d}t_0\,\mathrm{d}\vec{x}_0 K_1\left(t_1,\vec{x}_1;t_0,\vec{x}_0\right)\psi_0\left(t_0,\vec{x}_0\right)$$

6. Assume our initial wave function is given by a Gaussian test function:

$$\psi_1\left(t_1,\vec{x}_1\right) = \sqrt{\frac{m}{2\pi\hbar\varepsilon}}^4 \int \mathrm{d}t_0\,\mathrm{d}\vec{x}_0 \exp\left(-\frac{i\varepsilon}{\hbar}\frac{m}{2}\left(\left(\frac{t_1-t_0}{\varepsilon}\right)^2 - \left(\frac{\vec{x}_1-\vec{x}_0}{\varepsilon}\right)^2\right)\right)\sqrt[4]{\frac{1}{\pi\sigma^8}}\exp\left(-\frac{t_0^2}{2\sigma^2}-\frac{\left(\vec{x}_0-\langle\vec{x}_0\rangle\right)^2}{2\sigma^2}\right)$$

7. Again, the integrals converge step by step. As any square-integrable wave function may be written as a sum over such (see below) we have convergence. Of course to do this, we need covariant Morlet wavelets (see further below).

In many cases, an asymmetric treatment of time is harmless. But if we are analyzing time itself, then we do not want to wire the assumption that it is asymmetric into the maths. To do so would result in circular reasoning (see Price for an amusing discussion: (Price 1991)). The use of small imaginary factors/Euclidean time/Wick rotation will not work for analyses that are of time itself, as such approaches implicitly prejudge the conclusion.

### Morlet wavelets in one dimension

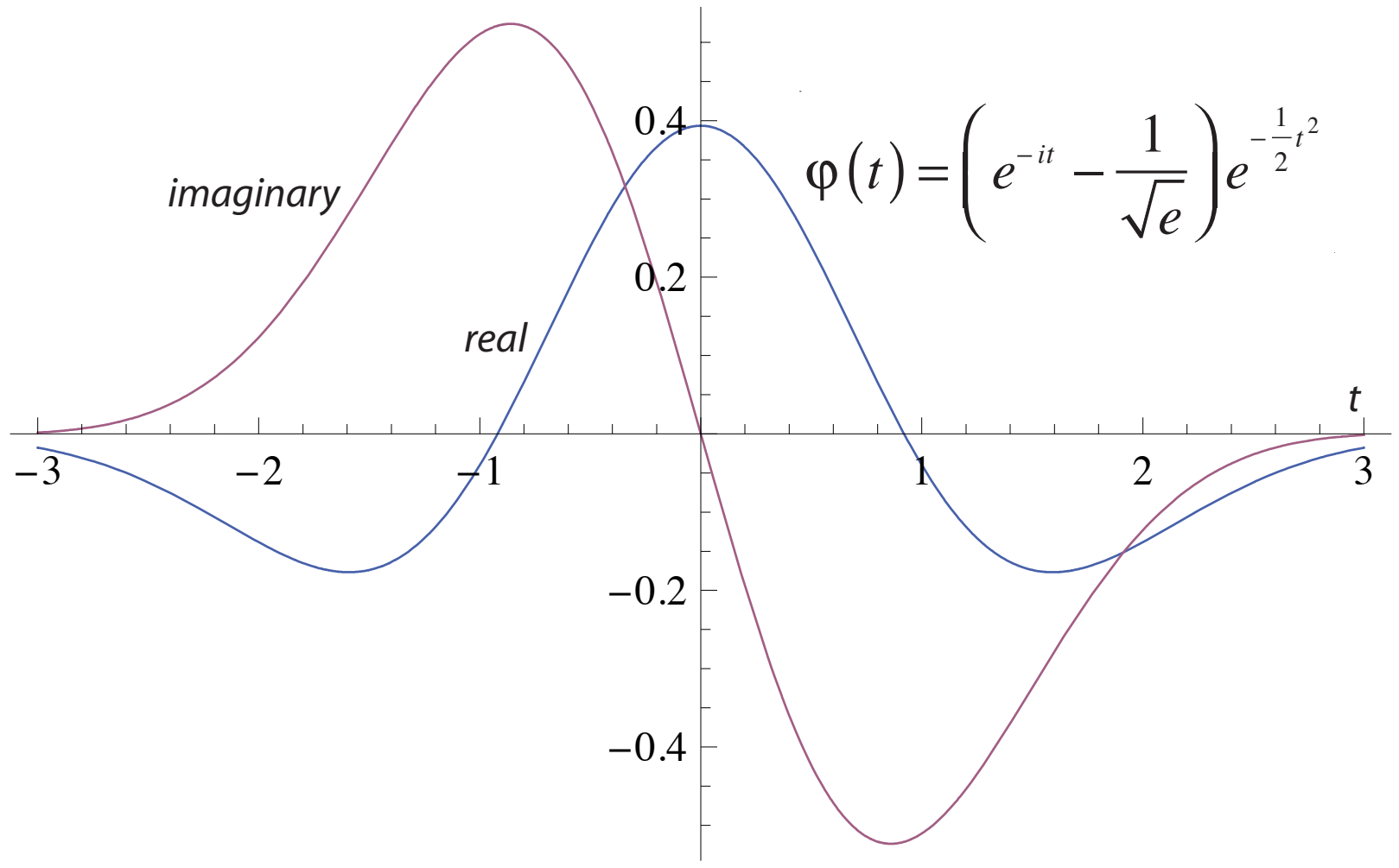

*Mother Morlet wavelet, with f = 1*

To generate a set of wavelets we start with a mother wavelet $\varphi(t)$. We get the general wavelet $\varphi_{sd}(t)$ by scaling the mother wavelet by a scale factor $s$ and displacing her by a displacement $d$:

$$\varphi_{sd}(t) \equiv \frac{1}{\sqrt{|s|}}\varphi\left(\frac{t-d}{s}\right).$$

As with life, so with wavelets: correct choice of your mother is essential for success. For Morlet wavelets the mother wavelet is given by:

$$\varphi(t) \equiv \left( e^{-ift} - e^{-\frac{f^2}{2}} \right) e^{-\frac{1}{2}t^2}$$

The second term is needed to satisfy the admissibility condition, discussed below. $t$ is often the time and $f$ may be thought of as a reference frequency.

As $f$ goes to infinity, the second term becomes less and less relevant – in many practical applications it is dropped (see for instance (Johnson 2009)). At the other extreme, when $f$ is zero, the mother wavelet is zero. We keep $f$ a variable to help in calculating the value of the admissibility constant $C_f$.

The general Morlet wavelet is created from the mother Morlet wavelet by scaling by $s$ and displacing by $d$:

$$\varphi_{sd}(t) \equiv \frac{1}{\sqrt{|s|}} \left( e^{-if\left(\frac{t-d}{s}\right)} - e^{-\frac{f^2}{2}} \right) e^{-\frac{1}{2}\left(\frac{t-d}{s}\right)^2}$$

Both scale and displacement run from -∞ to ∞. A negative scale just gives the complex conjugate of the Morlet wavelet with positive scale:

$$\varphi_{-s,d}(t) = \varphi^*_{s,d}(t)$$

The mother Morlet wavelet herself is given in this notation as the Morlet wavelet with scale factor one, displacement zero:

$$\varphi(t) = \varphi_{1,0}(t)$$

Any square integrable function $\psi$ may be expressed as a sum over Morlet wavelets. In principle this excludes $\delta$ functions and plane waves. We will see below that they are handled correctly however. The Morlet wavelet transform of a wave function $\psi$ (Morlet wavelet transform of $\psi$) is given by:

$$\tilde{\psi}_{sd} = \int_{-\infty}^{\infty} \mathrm{d}t \varphi^*_{sd}(t) \psi(t)$$

We get the original $\psi$ back (inverse Morlet wavelet transform of $\psi$) by integrating over the displacement and scale:

$$\psi(t) = \frac{1}{C_f} \int_{-\infty}^{\infty} \frac{\mathrm{d}s\mathrm{d}d}{s^2} \int_{-\infty}^{\infty} \mathrm{d}d \varphi_{sd}(t) \tilde{\psi}_{sd}$$

The admissibility constant is given by an integral over the square of the Fourier transform of the mother wavelet:

$$C_f \equiv 2\pi \int_{-\infty}^{\infty} \frac{\mathrm{d}w}{|w|} |\hat{\varphi}(w)|^2$$

In the general case we could use a different set of wavelets for the forward and the inverse transforms; it is one of the attractions of Morlet wavelets that we do not need to do this.

The wavelet decomposition fails if $C_f$ is not finite. For $C_f$ to be finite, we see we need the zero frequency component of the Fourier transform of the Morlet wavelet mother to be zero:

$$\hat{\varphi}(0) = 0 \Rightarrow \int_{-\infty}^{\infty} \mathrm{d}t \varphi(t) = 0.$$

The Fourier transform of the Morlet mother is real:

$$\hat{\varphi}(w) = (\exp(fw) - 1) \exp\left( -\frac{1}{2}f^2 - \frac{1}{2}w^2 \right)$$

The Fourier transform of the general Morlet wavelet is not:

$$\hat{\varphi}_{sd}(w) = \sqrt{|s|} e^{idw} (\exp(sfw) - 1) \exp\left( -\frac{1}{2}f^2 - \frac{1}{2}s^2 w^2 \right)$$

but may be written in terms of the Fourier transform of the mother:

$$\hat{\varphi}_{sd}(w) = \sqrt{|s|}e^{idw}\hat{\varphi}(sw).$$

*Normalization*

Morlet wavelets are not wave functions, but do not object to being treated as such. Their normalization is independent of their scale and displacement:

$$\int \mathrm{d}t \varphi_{sd}^{*}(t)\varphi_{sd}(t) = \left( e^{-f^2} - 2e^{-\frac{3f^2}{4}} + 1 \right)\sqrt{\pi}.$$

We can therefore write normalized Morlet wavelets as:

$$\varphi_{sd}^{(normalized)}(t) = \frac{1}{\sqrt[4]{\pi}\sqrt{e^{-f^2} - 2e^{-\frac{3f^2}{4}} + 1}}\varphi_{sd}(t).$$

*Resolution of unity*

We can establish the completeness of the wavelet transform by very general methods, see (Kaiser 1994). But if we are only concerned with Morlet wavelets, we can take advantage of their specific character to give a less general but more immediate proof[4].

If we substitute the integral for $\tilde{\psi}_{sd}$ in the integral for the inverse Morlet wavelet transform we get:

$$\psi(t) = \frac{1}{C_f}\int_{-\infty}^{\infty} \frac{\mathrm{d}s\mathrm{d}d\mathrm{d}t'}{s^2}\varphi_{sd}(t)\varphi_{sd}^{*}(t')\psi(t').$$

This will be true if we have[5]:

$$\delta(t-t') = \frac{1}{C_f}\int_{-\infty}^{\infty} \frac{\mathrm{d}s\mathrm{d}d}{s^2}\varphi_{sd}(t)\varphi_{sd}^{*}(t').$$

This looks like a familiar decomposition in terms of a set of states weighted by $\frac{1}{s^2}$. If we can show this directly, we will have shown we have a resolution of unity. To do this, we define the integral $I$:

$$I(t,t') \equiv \frac{1}{C_f}\int_{-\infty}^{\infty} \frac{\mathrm{d}s\mathrm{d}d}{s^2}\varphi_{sd}(t)\varphi_{sd}^{*}(t').$$

We wish to show that this integral gives the δ function. We write the Morlet wavelets in terms of their Fourier transforms to get:

$$I(t,t') = \frac{1}{C_f}\int_{-\infty}^{\infty} \frac{\mathrm{d}s\mathrm{d}d}{s^2}\int \frac{\mathrm{d}w}{\sqrt{2\pi}}\exp(-iwt)\hat{\varphi}_{sd}(w)\int \frac{\mathrm{d}w'}{\sqrt{2\pi}}\exp(iw't')\hat{\varphi}_{sd}^{*}(w').$$

Then we write the Fourier transforms of the wavelets in terms of the Fourier transform of the mother wavelet:

$$\frac{1}{C_f}\int_{-\infty}^{\infty} \frac{\mathrm{d}s\mathrm{d}d}{|s|}\int \frac{\mathrm{d}w}{\sqrt{2\pi}}\exp(-iw(t-d))\hat{\varphi}(sw)\int \frac{\mathrm{d}w'}{\sqrt{2\pi}}\exp(iw'(t'-d))\hat{\varphi}^{*}(sw').$$

We recognize the integral over $d$ as a δ function in $w$ and $w'$:

$$\int \frac{\mathrm{d}d}{2\pi}\exp(i(w-w')d) = \delta(w-w').$$

We use this hitherto disguised δ function to do the integral over $w'$:

---

[4] We are using the word "proof" in a relaxed rather than a rigorous sense.

[5] We are assuming we can freely interchange orders of integration.

$$I(t,t') = \frac{1}{C_f}\iint \frac{\mathrm{d}s\mathrm{d}w}{|s|} e^{-iwt} e^{iwt'} \hat{\varphi}(sw)\hat{\varphi}^*(sw)$$

We change the variable of integration to $s' = sw$ , then rename $s'$ to *s,* break up the integral into two parts, with scale *s* positive and *s* negative, and flip the sense of *s* in the negative *s* plane:

$$I(t,t') = \frac{1}{C_f}\int_0^\infty \frac{\mathrm{d}s}{s}\int_{-\infty}^\infty \mathrm{d}w e^{-iwt} e^{iwt'} \hat{\varphi}(s)\hat{\varphi}^*(s) + \frac{1}{C_f}\int_0^\infty \frac{\mathrm{d}s}{s}\int_{-\infty}^\infty \mathrm{d}w e^{-iwt} e^{iwt'} \hat{\varphi}(-s)\hat{\varphi}^*(-s)$$

We identify the *w* integration as another δ function, one which can come outside of the integrals:

$$I(t,t') = \delta(t-t')\frac{2\pi}{C_f}\int_0^\infty \frac{\mathrm{d}s}{s}\left(\hat{\varphi}(s)\hat{\varphi}^*(s) + \hat{\varphi}(-s)\hat{\varphi}^*(-s)\right)$$

We recognize the remaining integral as $C_f/2\pi$ so we have:

$$I(t,t') = \delta(t-t')$$

To show completeness we do not need the actual value of $C_f$, only that it is finite.

*Calculation of admissibility constant*

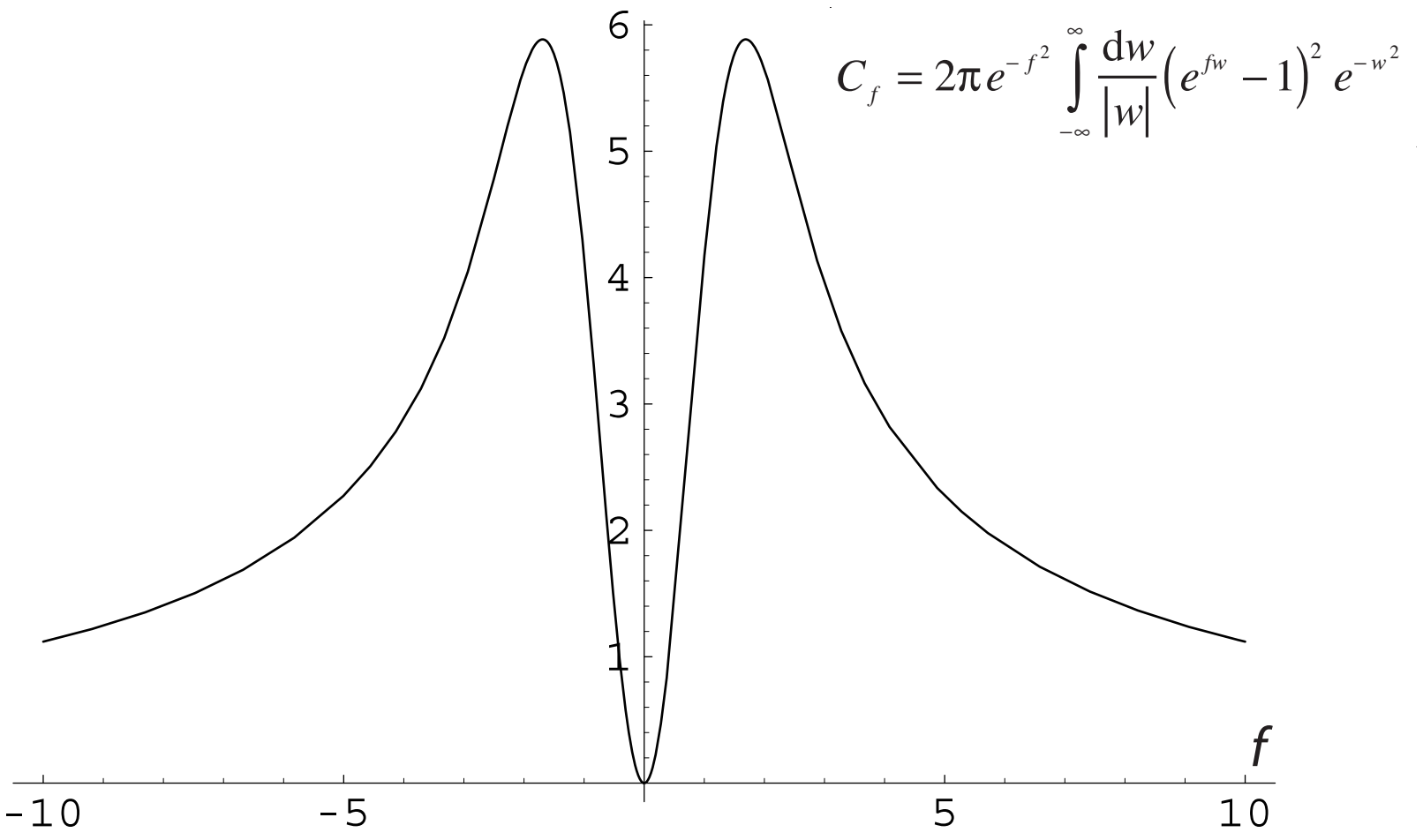

*Admissibility constant as a function of f*

Now, we compute the actual value of the admissibility constant. By substituting the explicit form of the Fourier transform of the mother Morlet wavelet in the formula for the admissibility constant we get:

$$C_f = 2\pi e^{-f^2}\int_{-\infty}^{\infty}\frac{\mathrm{d}w}{|w|}\left(e^{fw}-1\right)^2 e^{-w^2}$$

For convenience, we define a new integral *I*:

$$I(f) \equiv \int_{-\infty}^{\infty}\frac{\mathrm{d}w}{|w|}\left(e^{wf}-1\right)^2 e^{-w^2}$$

For *f* equal to zero, *I* is zero by inspection. This is expected given that the original mother wavelet is zero when *f* is zero. As *w* goes to zero, the integrand goes as $f^2 w$ so *I* is well-behaved in the small *w* limit. As *w* goes to ∞ the integral is obviously convergent; the exponential with argument quadratic in *w* ensures this. Therefore we can write *I* as:

$$I(f) = \int_0^f \mathrm{d}f' \frac{dI(f')}{df'}$$

The advantage of taking the derivative with respect to *f* is that it gets rid of the troubling factor of *w* in the denominator. The derivative of *I* with respect to *f* is:

$$\frac{dI(f)}{df} = 2\int_0^\infty dw\left(\left(e^{2wf} - e^{wf}\right) - \left(e^{-2wf} - e^{-wf}\right)\right)e^{-w^2}.$$

Or:

$$\frac{dI(f)}{df} = 4\int_0^\infty dw\left(\sinh(2wf) - \left(\sinh(wf)\right)\right)e^{-w^2}.$$

Giving:

$$\frac{dI(f)}{df} = 2\sqrt{\pi}\left(e^{f^2}erf(f) - e^{f^2/4}erf\left(\frac{f}{2}\right)\right).$$

We integrate this with respect to *f* to get a pair of hypergeometric functions:

$$I(f) = f^2\left(2\,{}_2F_2\left(1,1;\frac{3}{2},2;f^2\right) - {}_2F_2\left(1,1;\frac{3}{2},2;\frac{f^2}{4}\right)\right).$$

Therefore we have for $C_f$:

$$C_f = 2\pi e^{-f^2} f^2\left(2\,{}_2F_2\left(1,1;\frac{3}{2},2;f^2\right) - {}_2F_2\left(1,1;\frac{3}{2},2;\frac{f^2}{4}\right)\right).$$

The *F*'s are generalized hypergeometric functions. For *f* set to one we have:

$$C_1 = 2\pi e^{-1}\left(2\,{}_2F_2\left(1,1;\frac{3}{2},2;1\right) - {}_2F_2\left(1,1;\frac{3}{2},2;\frac{1}{4}\right)\right) \approx 4.1636.$$

This can be checked by doing the original integral numerically.
For small *f*, $C_f$ goes as:

$$\lim_{f\to 0} C_f \to 2\pi f^2.$$

For large *f*, $C_f$ goes as:

$$\lim_{f\to\infty} C_f \to \exp\left(-f^2\right).$$

At this point we have an explicit form for the inverse Morlet transform, so have reached our objective. We now apply the Morlet wavelet transform to some interesting cases.

### *Gaussian test functions*

Gaussian test functions (squeezed states) are the most important case:

$$\psi(t) = \sqrt[4]{\frac{1}{\pi\sigma^2}}\exp\left(-iE(t-\tau) - \frac{(t-\tau)^2}{2\sigma^2}\right).$$

The Fourier transform of a Gaussian test function is:

$$\hat{\psi}(w) = \sqrt[4]{\frac{1}{\pi}\sigma^2}\exp\left(iw\tau - \frac{(E-w)^2}{2}\sigma^2\right).$$

<u>Analysis</u>

The Morlet wavelet transform of a Gaussian test function is:

$$\tilde{\psi}_{sd} = \int_{-\infty}^{\infty} dt' \frac{1}{\sqrt{|s|}}\left(e^{if\frac{t'-d}{s}} - \exp\left(-\frac{f^2}{2}\right)\right)e^{-\frac{1}{2}\left(\frac{t'-d}{s}\right)^2}\sqrt[4]{\frac{1}{\pi\sigma^2}}\exp\left(-iE(t'-\tau) - \frac{(t'-\tau)^2}{2\sigma^2}\right).$$

The integral is elementary:

$$\tilde{\psi}_{sd} = \pi^{1/4}\sqrt{\frac{\sigma}{s^2+\sigma^2}}\sqrt{|s|}\sqrt{2}\left(e^{-i\frac{fs}{s^2+\sigma^2}(d-\tau)-\frac{1}{2}\frac{(Es-f)^2\sigma^2}{s^2+\sigma^2}} - e^{-\frac{f^2}{2}}e^{-\frac{1}{2}\frac{E^2s^2\sigma^2}{s^2+\sigma^2}}\right)e^{-iE\frac{\sigma^2}{s^2+\sigma^2}(d-\tau)-\frac{1}{2}\frac{(d-\tau)^2}{s^2+\sigma^2}}.$$

As expected, the leading term is greatest when the displacement $d = \tau$ and when the scale $s = \frac{f}{E}$. When $f$ is zero, the wavelet transform is zero.

## Inverse Morlet wavelet transform

As H. Dumpty might have put it, the tricky part isn't cutting the original wave function into parts, it's putting the parts back together again[6]. We expect the original $\psi$ will be given by:

$$\psi(t) = \frac{1}{C_f}\int_{-\infty}^{\infty}\frac{\mathrm{d}s\mathrm{d}d}{s^2}\varphi_{sd}(t)\tilde{\psi}_{sd}.$$

Without loss of generality, we simplify by assuming $\tau$ is zero:

$$\psi(t) = \frac{1}{C_f}\int_{-\infty}^{\infty}\frac{\mathrm{d}s\mathrm{d}d}{s^2}\left(\begin{array}{l}\pi^{1/4}\sqrt{\frac{\sigma}{s^2+\sigma^2}}\sqrt{2}\left(e^{-if\left(\frac{t-d}{s}\right)} - e^{-\frac{f^2}{2}}\right)e^{-\frac{1}{2}\left(\frac{t-d}{s}\right)^2} \\ \times\left(e^{-i\frac{fs}{s^2+\sigma^2}d-\frac{1}{2}\frac{(Es-f)^2\sigma^2}{s^2+\sigma^2}} - e^{-\frac{f^2}{2}}e^{-\frac{1}{2}\frac{E^2s^2\sigma^2}{s^2+\sigma^2}}\right)e^{-iE\frac{\sigma^2}{s^2+\sigma^2}d-\frac{1}{2}\frac{d^2}{s^2+\sigma^2}}\end{array}\right).$$

The integral over $d$ is straightforward, as all the terms are Gaussians in $d$:

$$\psi(t) = \frac{1}{C_f}\int_{-\infty}^{\infty}\frac{\mathrm{d}s}{|s|}2\pi^{3/4}\sqrt{\frac{\sigma}{2s^2+\sigma^2}}\left(\begin{array}{l}\exp\left(-\frac{2E^2s^2\sigma^2+2f^2(2s^2+\sigma^2)+2iE\sigma^2t+t^2}{2(2s^2+\sigma^2)}\right) \\ -2\exp\left(-\frac{3f^2s^2+2f^2\sigma^2-2Efs\sigma^2+2E^2s^2\sigma^2+2ifst+2iE\sigma^2t+t^2}{2(2s^2+\sigma^2)}\right) \\ +\exp\left(-\frac{2(f-Es)^2\sigma^2+2i(2fs+E\sigma^2)t+t^2}{2(2s^2+\sigma^2)}\right)\end{array}\right).$$

The limit as $s$ goes to zero is:

$$2\left(e^{-f^2-iEt-\frac{t^2}{2\sigma^2}}f^2\pi^{3/4}\left(\frac{1}{\sigma}\right)^{9/2}\left(\sigma^2+E^2\sigma^4-2iE\sigma^2t-t^2\right)\right)|s| + O\left[s^2\right].$$

The limit as $s$ goes to $\pm\infty$ is:

$$\pm\frac{\sqrt{2}e^{-f^2-\frac{\mathrm{E0}^2\sigma^2}{2}}\left(1-2e^{\frac{f^2}{4}}+e^{f^2}\right)\pi^{3/4}\sqrt{\sigma}}{s^2} + O\left[\frac{1}{s}\right]^3.$$

Our integral is therefore neither singular at the origin nor divergent at infinity. Of course, we expect this since we are guaranteed by the decomposition theorem that this integral will give the original Gaussian. To show explicitly we get the original Gaussian we take the Fourier transform of both sides, with respect to $t$. The simplification is dramatic and most of the factors come outside of the integral over $s$. On the right we get:

$$\frac{1}{C_f}2\pi\exp\left(-f^2\right)\sqrt[4]{\frac{\sigma^2}{\pi}}\exp\left(-\frac{(E-w)^2\sigma^2}{2}\right)\int_{-\infty}^{\infty}\frac{\mathrm{d}s}{|s|}e^{-s^2w^2}\left(-1+e^{fsw}\right)^2.$$

---

[6] For H. Dumpty's role in the analysis of the Stern-Gerlach experiment see Englert, B.-G., J. Schwinger, et al. (1988). "Is Spin Coherence Like Humpty-Dumpty? I. Simplified Treatment." Found. Phys. **18**: 1045--1056, Schwinger, J., M. O. Scully, et al. (1988). "Spin coherence and Humpty-Dumpty. II." Z. Phys. D **10**: 135, Scully, M. O., B.-G. Englert, et al. (1989). "Spin coherence and Humpty-Dumpty. III. The effects of observation." Phys. Rev. A **40**: 1775--1784.

We change variables $s' = sw$ and note the integral is essentially the admissibility constant. Factors cancel giving:

$$\sqrt[4]{\frac{\sigma^2}{\pi}}\exp\left(-\frac{(E-w)^2\sigma^2}{2}\right)\left(\frac{1}{C_f}2\pi\exp\left(-f^2\right)\int_{-\infty}^{\infty}\frac{\mathrm{d}s'}{|s'|}2e^{-s'^2}\left(-1+e^{fs'}\right)^2\right)=\sqrt[4]{\frac{\sigma^2}{\pi}}\exp\left(-\frac{(E-w)^2\sigma^2}{2}\right)=\hat{\psi}(w)$$

### *Other Applications*

We will compute the Morlet wavelet transforms of δ functions, plane waves, and – to achieve maximum self-referentiality – a Morlet wavelet itself.

#### δ functions

Since the δ function is not a square-integrable function, we are not guaranteed the wavelet transform will work. We therefore write the δ function as a limit of Gaussian test functions:

$$\delta(x)=\lim_{\sigma\to 0^+}\frac{1}{\sqrt{2\pi\sigma^2}}\exp\left(-\frac{x^2}{2\sigma^2}\right).$$

This lets us use the result for a Gaussian test function:

$$\tilde{\delta}_{sd}(\tau)=\lim_{\sigma\to 0^+}\sqrt{\frac{1}{s^2+\sigma^2}}\sqrt{|s|}\left(e^{-i\frac{fs}{s^2+\sigma^2}(d-\tau)-\frac{1}{2}\frac{f^2\sigma^2}{s^2+\sigma^2}}-e^{-\frac{f^2}{2}}\right)e^{-\frac{1}{2}\frac{(d-\tau)^2}{s^2+\sigma^2}}.$$

Taking the limit as σ goes to zero:

$$\tilde{\delta}_{sd}(\tau)=\frac{1}{\sqrt{|s|}}\left(e^{-if\left(\frac{d-\tau}{s}\right)}-e^{-\frac{f^2}{2}}\right)e^{-\frac{1}{2}\frac{(d-\tau)^2}{s^2}}.$$

This is itself a Morlet wavelet:

$$\tilde{\delta}_{sd}(\tau)=\varphi^*_{sd}(\tau).$$

We get the same result by computing the Morlet wavelet transform directly:

$$\tilde{\delta}_{sd}(\tau)=\int_{-\infty}^{\infty}\mathrm{d}t'\varphi^*_{sd}(t')\delta(t'-\tau)=\frac{1}{\sqrt{|s|}}\left(e^{if\frac{\tau-d}{s}}-\exp\left(-\frac{f^2}{2}\right)\right)e^{-\frac{1}{2}\left(\frac{\tau-d}{s}\right)^2}.$$

Since the demonstration of the resolution of unity only applies to square-integrable functions, we verify the inverse transform. We want to show:

$$\delta(t-\tau)=\frac{1}{C_f}\int_{-\infty}^{\infty}\frac{\mathrm{d}s\mathrm{d}d}{s^2}\varphi_{sd}(t)\varphi^{(*)}_{sd}(\tau).$$

However this is just what we showed when we computed the admissibility constant, so we are done.

#### Plane waves

The Morlet wave transform of a plane wave:

$$\phi(t)\equiv\frac{1}{\sqrt{2\pi}}\exp(-iEt)$$

is given by:

$$\tilde{\phi}_{sd}(E)=\int\mathrm{d}t\frac{1}{\sqrt{|s|}}\left(e^{if\frac{t-d}{s}}-\exp\left(-\frac{f^2}{2}\right)\right)e^{-\frac{1}{2}\left(\frac{t-d}{s}\right)^2}\frac{1}{\sqrt{2\pi}}\exp(-iEt).$$

We do the integral, discovering we get the Fourier transform of a Morlet wavelet:

$$\tilde{\phi}_{sd}(E)=\hat{\varphi}^{(*)}_{sd}(E).$$

For the inverse transform to be valid we require:

$$\phi(t)=\frac{1}{C_f}\int\frac{\mathrm{d}s\mathrm{d}d}{s^2}\varphi_{sd}(t)\hat{\varphi}_{sd}^{(*)}(E).$$

To show this, we take the Fourier transform of each side. On the left side we get:

$$\delta(w-E).$$

On the right side we have (writing the Fourier transforms of the Morlet wavelets in terms of the Fourier transforms of their mothers):

$$\frac{1}{C_f}\int\frac{\mathrm{d}s\mathrm{d}d}{s^2}\sqrt{|s|}\exp(idw)\hat{\varphi}(sw)\sqrt{|s|}\exp(-idE)\hat{\varphi}^*(sE).$$

The integral over $d$ is a δ function, which we pull out of the integral, leaving the now familiar admissibility constant behind:

$$\frac{2\pi}{C_f}\int\frac{\mathrm{d}s}{|s|}\hat{\varphi}(sw)\hat{\varphi}^*(sE)\delta(w-E)\rightarrow\frac{2\pi}{C_f}\int\frac{\mathrm{d}s}{|s|}\hat{\varphi}(sw)\hat{\varphi}^*(sw)\delta(w-E)=\delta(w-E).$$

<u>Morlet wavelet transform of a Morlet wavelet:</u>

We look at the Morlet wavelet transform of a Morlet wavelet. We use a normalized Morlet wavelet, with σ replacing $s$ and $E$ replacing $f$:

$$\psi_{\sigma E}(t)\equiv N_{\sigma E}\sqrt[4]{\frac{1}{\pi\sigma^2}}\left(e^{-iE(t-\tau)}-e^{-\frac{\sigma^2E^2}{2}}\right)e^{-\frac{1}{2}\left(\frac{t-\tau}{\sigma}\right)^2}$$

$$N_{\sigma E}\equiv\frac{\sqrt[4]{\pi}}{\sqrt{e^{-\sigma^2E^2}-2e^{-\frac{3\sigma^2E^2}{4}}+1}}.$$

The Morlet wavelet transform is given by:

$$\tilde{\psi}_{sd}^{(\sigma E)}=\int_{-\infty}^{\infty}\mathrm{d}t'\varphi_{sd}^*(t)\psi_{\sigma E}(t).$$

To apply the results for Gaussian test functions break out the incoming Morlet wavelet into its two Gaussian test functions; read off the results:

$$\tilde{\psi}_{sd}^{(\sigma E)}=\frac{\sqrt{2\pi\frac{\sigma}{s^2+\sigma^2}|s|}}{\sqrt{e^{-\sigma^2E^2}-2e^{-\frac{3\sigma^2E^2}{4}}+1}}\left(\begin{array}{l}\left(e^{-i\frac{fs}{s^2+\sigma^2}(d-\tau)-\frac{1}{2}\frac{(Es-f)^2\sigma^2}{s^2+\sigma^2}}-e^{-\frac{f^2}{2}}e^{-\frac{1}{2}\frac{E^2s^2\sigma^2}{s^2+\sigma^2}}\right)e^{-iE\frac{\sigma^2}{s^2+\sigma^2}(d-\tau)}\\-\left(e^{-i\frac{fs}{s^2+\sigma^2}(d-\tau)-\frac{1}{2}\frac{f^2\sigma^2}{s^2+\sigma^2}-\frac{\sigma^2E^2}{2}}-e^{-\frac{f^2}{2}-\frac{\sigma^2E^2}{2}}\right)\end{array}\right)e^{-\frac{1}{2}\frac{(d-\tau)^2}{s^2+\sigma^2}}.$$

## Covariant Morlet wavelets

### *Strategy*

We would like to generalize Morlet wavelets to four dimensions (one time, three space) in a way that is manifestly covariant. We will do this by taking the direct product of Morlet wavelets in time and the three space dimensions. The natural generalization of the Gaussian part of the one-dimensional Morlet wavelet is:

$$\exp\left(-\frac{x^2}{2}\right)\rightarrow\exp\left(-\frac{x_\mu x^\mu}{2}\right)=\exp\left(\frac{t^2}{2}-\frac{x^2+y^2+z^2}{2}\right).$$

This clearly diverges in $t$. We have to fix this without losing manifest covariance.

We will assume we start in a specific frame $M$. We will define the four-dimensional Morlet wavelet as the product of four one-dimensional Morlet wavelets, then write our results in a way that is Lorentz-invariant.

*Construction*

We take the four-dimensional mother Morlet wavelet as the direct product of four one-dimensional mother Morlet wavelets, one for each coordinate:

$$\varphi(t) \to \varphi(t)\varphi(x)\varphi(y)\varphi(x).$$

We write the mother Morlet wavelet out explicitly:

$$\varphi(t,x,y,z) = \left(\exp(-if_0 t) - \exp\left(-\frac{f_0^2}{2}\right)\right)\left(\exp(if_1 x) - \exp\left(-\frac{f_1^2}{2}\right)\right)\left(\exp(if_2 y) - \exp\left(-\frac{f_2^2}{2}\right)\right)\left(\exp(if_3 z) - \exp\left(-\frac{f_3^2}{2}\right)\right)$$
$$\times \exp\left(-\frac{t^2+x^2+y^2+z^2}{2}\right).$$

By scaling and displacing each component separately we get:

$$\varphi_{sd}(t,x,y,z) = \frac{1}{\sqrt{|s_0 s_1 s_2 s_3|}}\left(\exp\left(-if_0\frac{t-d_0}{s_0}\right) - \exp\left(-\frac{f_0^2}{2}\right)\right)\left(\exp\left(if_1\frac{x-d_1}{s_1}\right) - \exp\left(-\frac{f_1^2}{2}\right)\right)$$
$$\times\left(\exp\left(if_2\frac{y-d_2}{s_2}\right) - \exp\left(-\frac{f_2^2}{2}\right)\right)\left(\exp\left(if_3\frac{z-d_3}{s_3}\right) - \exp\left(-\frac{f_3^2}{2}\right)\right)$$
$$\times\exp\left(-\frac{1}{2}\left(\left(\frac{t-d_0}{s_0}\right)^2 + \left(\frac{x-d_1}{s_1}\right)^2 + \left(\frac{y-d_2}{s_2}\right)^2 + \left(\frac{z-d_3}{s_3}\right)^2\right)\right).$$

The Fourier transform of the mother Morlet wavelet is:

$$\hat{\varphi}(E,p_x,p_y,p_z) = (\exp(f_0 E)-1)(\exp(f_1 p_x)-1)(\exp(f_2 p_y)-1)(\exp(f_3 p_z)-1)$$
$$\times\exp\left(-\frac{f_0^2+f_1^2+f_2^2+f_3^2}{2} - \frac{E^2+p_x^2+p_y^2+p_z^2}{2}\right).$$

The Fourier transform of the general Morlet wavelet is:

$$\hat{\varphi}_{sd}(E,p_x,p_y,p_z) = \sqrt{|s_0 s_1 s_2 s_3|}\exp\left(i\left(d_0 E - d_1 p_x - d_2 p_y - d_3 p_z\right)\right)$$
$$\times(\exp(s_0 f_0 E)-1)(\exp(s_1 f_1 p_x)-1)(\exp(s_2 f_2 p_y)-1)(\exp(s_3 f_3 p_z)-1)$$
$$\times\exp\left(-\frac{f_0^2+f_1^2+f_2^2+f_3^2}{2} - \frac{s_0^2 E^2+s_1^2 p_x^2+s_2^2 p_y^2+s_3^2 p_z^2}{2}\right).$$

Now we have to promote various non-covariant bits to covariant bits.

The scale factors enter into the inverse Morlet integral in a slightly awkward way:

$$\int \frac{ds_0}{s_0^2}\frac{ds_1}{s_1^2}\frac{ds_2}{s_2^2}\frac{ds_3}{s_3^2}.$$

The simplest approach to this is to treat the four scale factors as so many scalars.

The obvious choices for the displacement *d* and the reference frequency *f* are to treat them as four vectors. For the displacement a single four vector will suffice:

$$d = (d_0, d_1, d_2, d_3).$$

We will need one four vector for each reference frequency:

$$F^{(0)} \equiv (f_0,0,0,0)$$
$$F^{(1)} \equiv (0,f_1,0,0)$$
$$F^{(2)} \equiv (0,0,f_2,0)$$
$$F^{(3)} \equiv (0,0,0,f_3).$$

For convenience, we define the sum over all four *F's* as:

$$F \equiv \sum_{n=0}^{3} F^{(n)} = \left(f_0, f_1, f_2, f_3\right).$$

This is also a four vector. Note that the raw frequencies $f_0, f_1, f_2, f_3$ are themselves scalars since they are defined with respect to the specific frame $M$.

To represent the sums as Lorentz invariants we define a set of second rank tensors (with their inverses):

$$\Sigma_{\mu}^{(n)\nu} \equiv \begin{pmatrix} s_0^{-n} & 0 & 0 & 0 \\ 0 & -s_1^{-n} & 0 & 0 \\ 0 & 0 & -s_2^{-n} & 0 \\ 0 & 0 & 0 & -s_3^{-n} \end{pmatrix}, \left(\frac{1}{\Sigma^{(n)}}\right)_{\mu}^{\nu} = \begin{pmatrix} s_0^{n} & 0 & 0 & 0 \\ 0 & -s_1^{n} & 0 & 0 \\ 0 & 0 & -s_2^{n} & 0 \\ 0 & 0 & 0 & -s_3^{n} \end{pmatrix}.$$

The first three – all we need – are:

$$\Sigma_{\mu}^{(0)\nu} \equiv \begin{pmatrix} 1 & 0 & 0 & 0 \\ 0 & -1 & 0 & 0 \\ 0 & 0 & -1 & 0 \\ 0 & 0 & 0 & -1 \end{pmatrix}, \Sigma_{\mu}^{(1)\nu} \equiv \begin{pmatrix} \frac{1}{s_0} & 0 & 0 & 0 \\ 0 & -\frac{1}{s_1} & 0 & 0 \\ 0 & 0 & -\frac{1}{s_2} & 0 \\ 0 & 0 & 0 & -\frac{1}{s_3} \end{pmatrix}, \Sigma_{\mu}^{(2)\nu} \equiv \begin{pmatrix} \frac{1}{s_0^2} & 0 & 0 & 0 \\ 0 & -\frac{1}{s_1^2} & 0 & 0 \\ 0 & 0 & -\frac{1}{s_2^2} & 0 \\ 0 & 0 & 0 & -\frac{1}{s_3^2} \end{pmatrix}.$$

By having the signature for each of these tensors be $(1,-1,-1,-1)$ we are ensuring that our Gaussian integrals will converge.

Written with these definitions the mother Morlet wavelet is:

$$\varphi\left(x_\mu\right) = \left(\prod_{n=0}^{3}\left(\exp\left(-iF^{(n)\mu}x_\mu\right) - \exp\left(-\frac{F^{(n)\mu}F_\mu}{2}\right)\right)\right)\exp\left(-x^\mu \frac{\Sigma_{\mu}^{(0)\nu}}{2} x_\nu\right)$$

and the general Morlet wavelet is:

$$\varphi_{\Sigma d}\left(x_\mu\right) = \sqrt{\det\left(\Sigma^{(1)}\right)}\left(\prod_{n=0}^{3}\left(\exp\left(-iF^{(n)\mu}\Sigma_{\mu}^{(0)\varpi}\Sigma_{\varpi}^{(1)\nu}\left(x_\nu - d_\nu\right)\right) - \exp\left(-\frac{F^{(n)\mu}\Sigma_{\mu}^{(0)\nu}F_{\nu}^{(n)}}{2}\right)\right)\right)\exp\left(-\left(x^\mu - d^\mu\right)\frac{\Sigma_{\mu}^{(2)\nu}}{2}\left(x_\nu - d_\nu\right)\right).$$

While we have worked this out in frame $M$, as it is written in terms of covariant quantities it is valid in all frames. We have therefore guaranteed Lorentz covariance of the Morlet wavelets.

Note that the choice of frame defines a set of Morlet wavelets; with each frame there is a distinct set of Morlet wavelets. If we have multiple frames we wish to work with we will need to tag each Morlet wavelet with the frame it comes from. Usually there is an obvious choice of frame, i.e. the center-of-mass frame.

With these definitions, the Fourier transform of the mother Morlet wavelet is:

$$\hat{\varphi}(p) = \left(\prod_{n=0}^{3}\left(\exp\left(F^{(n)}\frac{1}{\Sigma^{(0)}}p\right) - 1\right)\right)\exp\left(-F\frac{1}{2\Sigma^{(0)}}F - \frac{1}{2}p\frac{1}{2\Sigma^{(0)}}p\right).$$

The Fourier transform of the general Morlet wavelet is:

$$\hat{\varphi}_{\Sigma d}(p) = \sqrt{\frac{1}{\det\left(\Sigma^{(1)}\right)}}\exp(ipd)\left(\prod_{n=0}^{3}\left(\exp\left(F^{(n)}\frac{1}{\Sigma^{(1)}}p\right) - 1\right)\right)\exp\left(-F\frac{1}{2\Sigma^{(0)}}F - \frac{1}{2}p\frac{1}{2\Sigma^{(2)}}p\right).$$

### *Resolution of unity*

Any square integrable function $\psi(t,x,y,z)$ may be expressed as a sum over these Morlet wavelets. The covariant Morlet wavelet transform is given by:

$$\tilde{\psi}_{\Sigma d} = \int\limits_{-\infty}^{\infty} \mathrm{d}^4 x \varphi^*_{\Sigma d}(x) \psi(x)$$
.

And the inverse is given by:

$$\psi(x) = \frac{1}{C_{f_0} C_{f_1} C_{f_2} C_{f_3}} \int\limits_{-\infty}^{\infty} \left( \prod_{n=0}^{3} \frac{\mathrm{d}s_n}{|s_n|^2} \right) \int\limits_{-\infty}^{\infty} \mathrm{d}^4\, d \varphi_{\Sigma d}(x) \tilde{\psi}_{\Sigma d}$$
.

The resolution of unity and the values of the constants of admissibility follow directly from the results for one dimension.

The solutions for Gaussian test functions, δ functions, and plane waves are merely the direct products of the corresponding one-dimensional wave functions.

We have therefore reached our goal: to generalize the Morlet wavelet transform to four dimensions in a way which is manifestly covariant.

### *Alternative approaches*

Alternative (and more sophisticated) lines of attack are possible. For instance in (Antoine, Murenzi et al. 2004) or in (Perel and Sidorenko 2007) two dimensional wavelets are generated from the mother wavelet by using displacements, rotations (in the x-y plane), and a single scale factor:

$$\varphi_{Rds}(x,y) = \frac{1}{s} \varphi^{(mom)} \left( \vec{R} \cdot \frac{(\vec{r} - \vec{d})}{s} \right)$$

where $R$ is a rotation matrix (in two dimensions).

By analogy, we could generalize one-dimensional wavelets to four dimensions by using displacements $d$, Lorentz transformations $\Lambda$, and a single dilation $s$:

$$\varphi_{\Lambda ds}(x_\mu) = \frac{1}{s^2} \varphi^{(mom)} \left( \frac{1}{s} \Lambda_\mu^\nu (x_\nu - d_\nu) \right)$$
.

But establishing convergence, verifying the resolution of unity, and computing the admissibility constant for these wavelets would be a new project. Our immediate requirement is merely to establish that there is at least *one* set of covariant Morlet wavelets.

### Summary

The naive use of plane wave/δ function decomposition can create artificial difficulties in the analysis of foundational questions of quantum mechanics[7].

The use of Morlet wavelet decomposition avoids these difficulties.

With the explicit calculation of the admissibility constant and the demonstration of covariant Morlet wavelets, we have eliminated two of the barriers to full use of this powerful technology for the analysis of foundational questions in quantum mechanics.

## Conventions for the Fourier transform

For the Fourier transform from time to frequency we are using:

$$\hat{f}(w) \equiv \frac{1}{\sqrt{2\pi}} \int\limits_{-\infty}^{\infty} \mathrm{d}t \exp(iwt) f(t)$$

with inverse Fourier transform:

$$f(t) = \frac{1}{\sqrt{2\pi}} \int\limits_{-\infty}^{\infty} \mathrm{d}w \exp(-iwt) \hat{f}(w).$$

In the Fourier transform in four dimensions time and space enter with opposite signs:

[7] That the use of unlocalized wave functions creates artificial difficulties suggests that the localization of the wave function in time and frequency/space and momentum should be regarded as a fundamental attribute.

$$\hat{f}\left(w,\vec{k}\right) \equiv \frac{1}{4\pi^2}\int_{-\infty}^{\infty} \mathrm{d}t\,\mathrm{d}\vec{x}\exp\left(iwt - i\vec{k}\cdot\vec{x}\right) f\left(t,\vec{x}\right)$$

with inverse Fourier transform:

$$f\left(t,\vec{x}\right) = \frac{1}{4\pi^2}\int_{-\infty}^{\infty} \mathrm{d}w\,\mathrm{d}\vec{k}\exp\left(-iwt + i\vec{k}\cdot\vec{x}\right)\hat{f}\left(w,\vec{k}\right).$$